\documentclass[final]{siamltex}
\usepackage{graphicx,amssymb,psfrag}

\newcommand{\BEAS}{\begin{eqnarray*}}
\newcommand{\EEAS}{\end{eqnarray*}}
\newcommand{\BEA}{\begin{eqnarray}}
\newcommand{\EEA}{\end{eqnarray}}
\newcommand{\BEQ}{\begin{equation}}
\newcommand{\EEQ}{\end{equation}}
\newcommand{\BIT}{\begin{itemize}}
\newcommand{\EIT}{\end{itemize}}
\newcommand{\BNUM}{\begin{enumerate}}
\newcommand{\ENUM}{\end{enumerate}}

\newcommand{\BA}{\begin{array}}
\newcommand{\EA}{\end{array}}

\newcommand{\ie}{{\it i.e.}}

\newcommand{\ones}{\mathbf 1}

\newcommand{\reals}{{\mbox{\bf R}}}

\newcommand{\symm}{{\mbox{\bf S}}}  

\newcommand{\Rank}{\mathop{\bf Rank}}
\newcommand{\Card}{\mathop{\bf Card}}
\newcommand{\Tr}{\mathop{\bf Tr}}
\newcommand{\lambdamax}{{\lambda_{\rm max}}}

\title{A direct formulation for sparse PCA using semidefinite programming\thanks{A preliminary version of this paper appeared in the proceedings of the Neural Information Processing Systems (NIPS) 2004 conference and the associated preprint is on arXiv as cs.CE/0406021.}}

\begin{document}

\author{Alexandre d'Aspremont\thanks{ORFE Dept., Princeton University, Princeton, NJ 08544. \texttt{aspremon@princeton.edu}}\and Laurent El Ghaoui\thanks{EECS  Dept., U.C. Berkeley, Berkeley, CA 94720. \texttt{elghaoui@eecs.berkeley.edu}}\and Michael I. Jordan\thanks{EECS and  Statistics Depts., U.C. Berkeley, Berkeley, CA 94720. \texttt{jordan@cs.berkeley.edu}}\and Gert R. G.  Lanckriet\thanks{ECE Dept., U.C. San Diego, La Jolla, CA 92093. \texttt{gert@ece.ucsd.edu}} }

\maketitle

\begin{abstract}
Given a covariance matrix, we consider the problem of maximizing the variance explained by a particular linear combination of the input variables while constraining the number of nonzero coefficients in this combination. This problem arises in the decomposition of a covariance matrix into sparse factors or sparse PCA, and has wide applications ranging from biology to finance.  We use a modification of the classical variational representation of the largest eigenvalue of a symmetric matrix, where cardinality is constrained, and derive a semidefinite programming based relaxation for our problem. We also discuss Nesterov's smooth minimization technique applied to the semidefinite program arising in the semidefinite relaxation of the sparse PCA problem.  The method has complexity $O(n^4 \sqrt{\log(n)}/\epsilon)$, where $n$ is the size of the underlying covariance matrix, and $\epsilon$ is the desired absolute accuracy on the optimal value of the problem.
\end{abstract}



\pagestyle{myheadings}
\thispagestyle{plain}
\markboth{A. D'ASPREMONT, L. EL GHAOUI, M. I. JORDAN AND G.R.G. LANCKRIET}{SPARSE PCA USING SEMIDEFINITE PROGRAMMING}


\section{Introduction}
\label{sec:intro} Principal component analysis (PCA) is a popular tool for data analysis, data compression and data visualization. It has applications throughout science and engineering.  In essence, PCA finds linear combinations of the variables (the so-called principal components) that correspond to directions of maximal variance in the data. It can be performed via a singular value decomposition (SVD) of the data matrix $A$, or via an eigenvalue decomposition if $A$ is a covariance matrix.

The importance of PCA is due to several factors. First, by capturing directions of maximum variance in the data, the principal components offer a way to compress the data with minimum information loss. Second, the principal components are
uncorrelated, which can aid with interpretation or subsequent statistical analysis. On the other hand, PCA has a number of well-documented disadvantages as well.  A particular disadvantage that is our focus here is the fact that the principal components 
are usually linear combinations of \emph{all} variables. That is, all weights in the linear combination (known as \emph{loadings}) are typically non-zero. In many applications, however, the coordinate axes have a physical interpretation; in biology for example, each axis might correspond to a specific gene.  In these cases, the interpretation of the principal components would be facilitated if these components involved very few non-zero loadings (coordinates).  Moreover, in certain applications, e.g., financial asset trading strategies based on principal component techniques, the sparsity of the loadings has important consequences, since fewer non-zero loadings imply fewer  transaction costs.

It would thus be of interest to discover \emph{sparse principal components}, \ie,~sets of sparse vectors spanning a low-dimensional space that explain most of the variance present in the data. To achieve this, it is necessary to sacrifice some of the explained variance and the orthogonality of the principal components, albeit hopefully not too much.

Rotation techniques are often used to improve interpretation of
the standard principal components (see \cite{Joll95} for example). Vines or Kolda and O'Leary \cite{Vine00,Kold00} considered simple principal components by restricting the loadings to take values from a small set of allowable integers, such as $0$, $1$, and $-1$. Cadima and Jolliffe \cite{cadi95} proposed an ad hoc way to deal with the problem, where the loadings with small absolute value are thresholded to zero. We will call this approach ``simple thresholding.'' Later, algorithms such as SCoTLASS \cite{Joll03} and SLRA \cite{Zhan02a,Zhan04} were introduced to find modified principal components with possible zero loadings. Finally, Zou, Hastie and Tibshirani \cite{Zou04} propose a new approach called sparse PCA (SPCA) to find modified components with zero loading in very large problems, by writing PCA as a regression-type optimization problem. This allows the application of LASSO \cite{tibs96}, a penalization technique based on the $l_1$ norm. All these methods are either significantly suboptimal (thresholding) or nonconvex (SCoTLASS, SLRA, SPCA).

In this paper, we propose a direct approach (called DSPCA in what follows) that improves the sparsity of the principal components by directly incorporating a sparsity criterion in the PCA problem formulation, then forming a \emph{convex relaxation} of the problem. This convex relaxation turns out to be a semidefinite program.

Semidefinite programs (SDP) can be solved in polynomial time via general-purpose interior-point methods \cite{stur99,Toh96}. This suffices for an initial empirical study of the properties of DSPCA and for comparison to the algorithms discussed above on small problems. For high-dimensional problems, however, the general-purpose methods are not viable and it is necessary to exploit problem structure. Our particular problem can be expressed as a saddle-point problem which is well suited to recent algorithms based on a \emph{smoothing} argument combined with an optimal first-order smooth minimization algorithm~\cite{Nest03,Nemi04,BenT04}. These algorithms offer a significant reduction in computational time compared to generic interior-point SDP solvers. This also represents a change in the \emph{granularity} of the solver, requiring a larger number of significantly cheaper iterations. In many practical problems this is a desirable tradeoff; interior-point solvers often run out of memory in the first iteration due to the necessity of forming and solving large linear systems.  The lower per-iteration memory requirements of the first-order algorithm described here means that considerably larger problems can be solved, albeit more slowly.

This paper is organized as follows.  In section  \ref{sec:sparse-eig}, we show how to efficiently maximize the variance of a projection while constraining the cardinality (number of nonzero coefficients) of the vector defining the projection.  We achieve this via a semidefinite relaxation. We briefly explain how to generalize this approach to non-square matrices and formulate a robustness interpretation of our technique. We also show how this interpretation can be used in the decomposition of a matrix into sparse factors. Section \ref{sec:large-scale} describes how Nesterov's smoothing technique (see \cite{Nest03}, \cite{Nest04a}) can be used to solve large problem instances efficiently. Finally, section \ref{sec:numerical-results} presents applications and numerical experiments comparing our method with existing techniques.

\subsection*{Notation}
In this paper, $\symm^n$ is the set of symmetric matrices of size
$n$, and $\Delta_n$ the spectahedron (set of positive semidefinite matrices with unit trace). We denote by
$\ones$ a vector of ones, while $\Card (x)$ denotes the cardinality (number of non-zero elements) of a vector $x$ and  $\Card (X)$ is the number of non-zero coefficients in the matrix $X$. For $X\in\symm^n$, $X \succeq 0$ means that $X$ is positive semidefinite, $\|X\|_F$ is the Frobenius norm of $X$, i.e., $\|X\|_F=\sqrt{\Tr(X^2)}$, $\lambda^{\mathrm{max}}(X)$ is the maximum eigenvalue of $X$ and $\|X\|_\infty = \max_{\{i,j=1,\ldots,n\}} |X_{ij}|$, while $|X|$ is the matrix whose elements are the absolute values of the elements of $X$. Finally, for matrices $X,Y\in \symm^n$, $X \circ Y$ is the Hadamard (componentwise or Schur) product of $X$ and $Y$.

\section{Semidefinite Relaxation}
\label{sec:sparse-eig} In this section, we derive a semidefinite programming (SDP) relaxation for the problem of maximizing the variance explained by a vector while constraining its cardinality. We formulate this as a variational problem, then obtain a lower bound on its optimal value via an SDP relaxation (we refer the reader to \cite{Vand96} or \cite{Boyd03} for an overview of semidefinite programming).

\subsection{Sparse Variance Maximization}
Let $A\in \symm^n$ be a given symmetric matrix and $k$ be an integer with $1 \leq k \leq n$. Given the matrix $A$ and assuming without loss of generality that $A$ is a covariance matrix (i.e. $A$ is positive semidefinite), we consider the problem of maximizing the variance of a vector $x\in\reals^n$ while constraining its cardinality:
\BEQ
\label{eq:variat-prog}
\BA{ll}
\mbox{maximize} & x^TAx\\
\mbox{subject to} & \|x\|_2=1\\
& \Card(x) \leq k.
\EA
\EEQ
Because of the cardinality constraint, this problem is hard (in fact, NP-hard) and we look here for a convex, hence efficient, relaxation. 
\subsection{Semidefinite Relaxation}
Following the \emph{lifting procedure} for semidefinite relaxation described in \cite{Lova91}, \cite{Aliz95}, \cite{Lema99} for example, we rewrite (\ref{eq:variat-prog}) as:
\BEQ \label{eq-variat-matrix-prog}
\BA{ll}
\mbox{maximize} & \Tr(AX)\\
\mbox{subject to} & \Tr(X)=1\\
& \Card(X) \leq k^2\\
& X \succeq 0,~\Rank(X)=1,
\EA \EEQ
in the (matrix) variable $X\in \symm^n$. Programs (\ref{eq:variat-prog}) and (\ref{eq-variat-matrix-prog}) are equivalent, indeed if $X$ is a solution to the above problem, then $X \succeq 0$ and $\Rank(X)=1$ mean that we have $X=xx^T$, while $\Tr(X)=1$ implies that $\|x\|_2=1$. Finally, if $X=xx^T$ then $\Card(X) \leq k^2$ is equivalent to $\Card(x) \leq k$. We have made some progress by turning the convex maximization objective $x^TAx$ and the nonconvex constraint $\|x\|_2=1$ into a linear constraint and linear objective. Problem (\ref{eq-variat-matrix-prog}) is, however, still nonconvex and we need to relax both the rank and cardinality constraints. 

Since for every $u\in\reals^n$, $\Card(u)=q$ implies $\|u\|_1 \leq \sqrt{q} \|u\|_2$, we can replace the nonconvex constraint $\Card(X) \leq k^2$, by a weaker but convex constraint: $\ones ^T |X| \ones \leq k$, where we exploit the property that $\|X\|_F= \sqrt{x^Tx}=1$ when $X=xx^T$ and $\Tr(X)=1$. If we drop the rank constraint, we can form a relaxation of  (\ref{eq-variat-matrix-prog}) and (\ref{eq:variat-prog}) as:
\BEQ \label{eq-variat-relax} \BA{lll}
\mbox{maximize} & \Tr(AX)\\
\mbox{subject to} & \Tr(X)=1\\
& \ones ^T |X| \ones \leq k\\
& X \succeq 0,
\EA \EEQ
which is a semidefinite program in the variable $X\in\symm^n$, where $k$ is an integer parameter controlling the sparsity of the solution. The optimal value of this program will be an upper bound on the optimal value of the variational problem in (\ref{eq:variat-prog}). Here, the relaxation of $Card(X)$ in $\ones ^T |X| \ones$ corresponds to a classic technique which replaces the (non-convex) cardinality or $l_0$ norm of a vector $x$ with its largest convex lower bound on the unit box: $|x|$, the $l_1$ norm of $x$ (see \cite{Boyd00} or \cite{Dono05} for other applications).

\subsection{Extension to the Non-Square Case}
\label{ss:sparse-svd}
Similar reasoning applies to the case of a non-square matrix $A\in\reals^{m \times n}$, and the problem:
\[
\BA{ll}
\mbox{maximize} & u^TAv \\
\mbox{subject to} &\|u\|_2 = \|v\|_2 = 1\\
&\Card(u) \leq k_1,~\Card(v) \leq k_2,
\EA
\]
in the variables $(u,v) \in \reals^m \times \reals^n$ where $k_1 \leq m$, $k_2 \leq n$ are fixed integers controlling the sparsity. This can be relaxed to:
\[
\BA{ll}
\mbox{maximize} & \Tr (A^TX^{12}) \\
\mbox{subject to} & X \succeq 0,~\Tr (X^{ii}) = 1\\
& \ones^T|X^{ii}|\ones \leq k_i,\quad i=1,2 \\
& \ones^T|X^{12}|\ones \leq \sqrt{k_1k_2},
\EA
\]
in the variable $X\in\symm^{m+n}$ with blocks $X^{ij}$ for
$i,j=1,2$. Of course, we can consider several variations on this, such as constraining $\Card(u,v)=\Card(u)+\Card(v)$.

\section{A Robustness Interpretation}
\label{sec:robust} In this section, we show that problem (\ref{eq-variat-relax}) can be interpreted as a robust formulation of the maximum eigenvalue problem, with additive, componentwise uncertainty in the input matrix $A$.  We again assume $A$ to be symmetric and positive semidefinite.

In the previous section, we considered a cardinality-constrained variational formulation of the maximum eigenvalue problem:
\[
\BA{ll}
\mbox{maximize} & x^TAx\\
\mbox{subject to} & \|x\|_2=1\\
& \Card(x) \leq k.
\EA
\]
Here we look instead at a variation in which we penalize the
cardinality and solve:
\BEQ
\BA{ll}
\mbox{maximize} & x^TAx - \rho \Card^2(x)\\
\mbox{subject to} & \|x\|_2=1,\\
\EA
\EEQ
in the variable $x\in\reals^n$, where the parameter $\rho>0$
controls the magnitude of the penalty. This problem is again nonconvex and very difficult to solve. As in the last section, we can form the equivalent program:
\[
\BA{ll}
\mbox{maximize} & \Tr(AX) - \rho \Card(X)\\
\mbox{subject to} & \Tr(X)=1\\
& X \succeq 0,~\Rank(X)=1,
\EA
\]
in the variable $X\in \symm^n$. Again, we get a relaxation of this program by forming:
\BEQ
\label{eq:penalized-relax}
\BA{ll}
\mbox{maximize} & \Tr(AX) - \rho \ones^T |X| \ones\\
\mbox{subject to} & \Tr(X)=1\\
& X \succeq 0,
\EA \EEQ
which is a semidefinite program in the variable $X\in \symm^n$,
where $\rho>0$ controls the magnitude of the penalty. We can rewrite this problem as:
\BEQ
\label{eq:saddle}
\max_{X \succeq 0, \Tr(X)=1} ~ \min_{|U_{ij}|\leq \rho} \Tr(X(A+U)) 
\EEQ
in the variables $X\in\symm^n$ and $U\in\symm^n$. This yields the following dual to (\ref{eq:penalized-relax}):
\BEQ
\label{eq:dual-robust}
\BA{ll}
\mbox{minimize} & \lambda^{\mathrm{max}}(A+U)\\
\mbox{subject to} & |U_{ij}|\leq \rho,\quad i,j=1,\ldots,n,
\EA \EEQ
which is a maximum eigenvalue problem with variable $U\in
\symm^n$. This gives a natural robustness interpretation to the relaxation in (\ref{eq:penalized-relax}): it corresponds to a worst-case maximum eigenvalue computation, with componentwise bounded noise of intensity $\rho$ imposed on the matrix coefficients.

Let us first remark that $\rho$ in (\ref{eq:penalized-relax}) corresponds to the optimal Lagrange multiplier in (\ref{eq-variat-relax}). Also, the KKT conditions (see \cite[\S5.9.2]{Boyd03}) for problem (\ref{eq:penalized-relax}) and (\ref{eq:dual-robust}) are given by:
\BEQ
\label{eq:KKT}
\left\{
\BA{l}
(A+U)X=\lambda^{\mathrm{max}}(A+U)X\\
U\circ X=\rho |X|\\
\Tr(X)=1,~X \succeq 0\\
|U_{ij}|\leq \rho,\quad i,j=1,\ldots,n.\\
\EA 
\right.
\EEQ
If the eigenvalue $\lambda^{\mathrm{max}}(A+U)$ is simple (when, for example, $\lambda^{\mathrm{max}}(A)$ is simple and $\rho$ is sufficiently small), the first condition means that $\Rank(X)=1$ and the semidefinite relaxation is \emph{tight}, with in particular $Card(X)=Card(x)^2$ if $x$ is the dominant eigenvector of $X$. When the optimal solution $X$ is not of rank one because of degeneracy (i.e. when $\lambda^{\mathrm{max}}(A+U)$ has multiplicity strictly larger than one), we can truncate $X$ as in \cite{Aliz95,Lema99}, retaining only the dominant eigenvector $x$ as an approximate solution to the original problem. In that degenerate scenario however, the dominant eigenvector of $X$ is not guaranteed to be as sparse as the matrix itself.

\section{Sparse Decomposition} \label{sec:sparse-dec} Using the results obtained in the previous two sections we obtain a sparse equivalent to the PCA decomposition. Given a matrix $A_1\in\symm^n$, our objective is to decompose it in factors with target sparsity $k$. We solve the relaxed problem in (\ref{eq-variat-relax}):
\BEQ
\label{eq:sparse-decomp}
\BA{ll}
\mbox{maximize} & \Tr(A_1X)\\
\mbox{subject to} & \Tr(X)=1\\
 & \ones ^T |X| \ones \leq k\\
 & X \succeq 0.
\EA
\EEQ
Letting $X_1$ denote the solution, we truncate $X_1$, retaining only the dominant (sparse) eigenvector $x_1$. Finally, we deflate $A_1$ to obtain
\[
A_2=A_1-(x_1^TA_1x_1)x_1x_1^T,
\]
and iterate to obtain further components. The question is now: When do we stop the decomposition? 

In the PCA case, the decomposition stops naturally after $\Rank(A)$ factors have been found, since $A_{\Rank(A)+1}$ is then equal to zero. In the case of the sparse decomposition, we have no guarantee that this will happen. 
Of course, we can add an additional set of linear constraints $x_i^TXx_i=0$ to problem (\ref{eq:sparse-decomp}) to explicitly enforce the orthogonality of $x_1,\ldots,x_n$ and the decomposition will then stop after a maximum of $n$ iterations.
Alternatively, the robustness interpretation gives us a natural stopping criterion: if all the coefficients in $|A_{i}|$ are smaller than the noise level $\rho^{\star}$ (computed in the last section) then we must stop since the matrix is essentially indistinguishable from zero. Thus, even though we have no guarantee that the algorithm will terminate with a zero matrix, in practice the decomposition will terminate as soon as the coefficients in $A$ become indistinguishable from the noise.

\section{Algorithm} \label{sec:large-scale}
For small problems, the semidefinite program (\ref{eq:sparse-decomp}) can be solved efficiently using interior-point solvers such as SEDUMI \cite{stur99} or SDPT3 \cite{Toh96}. For larger-scale problems, we need to resort to other types of convex optimization algorithms because the $O(n^2)$ constraints implicitly contained in $\ones ^T |X| \ones \leq k$ make the memory requirements of Newton's method prohibitive. Of special interest are the algorithms recently presented in \cite{Nest03,Nemi04,BenT04}. These are first-order methods specialized to problems such as (\ref{eq:saddle}) having a specific saddle-point structure. These methods have a significantly smaller memory cost per iteration than interior-point methods and enable us to solve much larger problems. Of course, there is a price: for {\em fixed} problem size, the first-order methods mentioned above converge in $O(1/\epsilon)$ iterations, where $\epsilon$ is the required accuracy on the optimal value, while interior-point methods converge as $O(\log(1/\epsilon))$. Since the problem under consideration here does not require a high degree of precision, this slow convergence is not a major concern. In what follows, we adapt the algorithm in \cite{Nest03} to our particular constrained eigenvalue problem.

\subsection{A Smoothing Technique}
The numerical difficulties arising in large scale semidefinite programs stem from two distinct origins. First, there is an issue of \emph{memory}: beyond a certain problem size $n$, it becomes essentially impossible to form and store any second order information (Hessian) on the problem, which is the key to the numerical efficiency of interior-point SDP solvers. Second, \emph{smoothness} is an issue: the constraint $X\succeq 0$ is not smooth, hence the number of iterations required to solve problem (\ref{eq-variat-relax}) using first-order methods such as the bundle code of \cite{helm00} (which do not form the Hessian) to an accuracy $\epsilon$ is given by $O(1/\epsilon^2)$. In general, this complexity bound is tight and cannot be improved without additional structural information on the problem. Fortunately, in our case we do have structural information available that can be used to bring the complexity down from $O(1/\epsilon^2)$ to $O(1/\epsilon)$. Furthermore, the cost of each iteration is equivalent to that of computing a matrix exponential (roughly $O(n^3)$).

Recently, \cite{Nest03} and \cite{Nest04a} (see also \cite{Nemi04}) proposed an efficient first-order scheme for convex minimization based on a smoothing argument. The main structural assumption on the function to minimize is that it has a \emph{saddle-function} format:
\BEQ
\label{eq:sadf} f(x)=\hat f(x) + \max_u \{\langle Tx,u\rangle-\hat
\phi(u)~:~u\in Q_2\}
\EEQ
where $f$ is defined over a compact convex set $Q_1\subset\reals^n$, $\hat f(x)$ is convex and differentiable and has a Lipschitz continuous gradient with constant $M\geq 0$, $T$ is an element of $\reals^{n \times n}$ and $\hat\phi(u)$ is a continuous convex function over some closed compact set $Q_2\subset \reals^n$. This assumes that the function $\hat\phi(u)$ and the set $Q_2$ are simple enough so that the optimization subproblem in $u$ can be solved very efficiently. When a function $f$ can be written in this particular format, \cite{Nest03} uses a \emph{smoothing technique} to show that the complexity (number of iterations required to obtain a solution with absolute precision $\epsilon$) of solving:
\BEQ
\label{eq:sad-min} \min_{x\in Q_1} f(x)
\EEQ
falls from $O(1/\epsilon^2)$ to $O(1/\epsilon)$. This is done in two steps.
\paragraph{\bf Regularization}
By adding a strongly convex penalty to the saddle function representation of $f$ in (\ref{eq:sadf}), the algorithm first computes a \emph{smooth} $\epsilon$-approximation of $f$ with Lipschitz continuous gradient. This can be seen as a generalized Moreau-Yosida regularization step (see \cite{Lema97} for example).
\paragraph{\bf Optimal first-order minimization} 
The algorithm then applies the optimal first-order scheme for functions with Lipschitz continuous gradient detailed in \cite{Nest83} to the regularized function. Each iteration requires an efficient computation of the regularized function value and its gradient. As we will see, this can be done explicitly in our case, with a complexity of $O(n^3)$ and memory requirements of $O(n^2)$.

\subsection{Application to Sparse PCA}
Given a symmetric matrix $A\in\symm^n$, we consider the problem given in (\ref{eq:saddle}) (where we can assume without loss of generality that $\rho=1$):
\BEQ\label{eq:orig-pb}
\BA{ll}
\mbox{maximize} & \Tr(AX) - \ones^T |X| \ones\\
\mbox{subject to} & \Tr(X)=1\\
& X \succeq 0,
\EA
\EEQ
in the variable $X\in\symm^n$. Duality allows us to rewrite this in the saddle-function format:
\BEQ
\min_{U \in {\cal Q}_1} \: f(U),
\EEQ
where
\[\BA{cc}
{\cal Q}_1 = \left\{ U \in \symm^n :|U_{ij}|\leq1,~i,j=1,\ldots,n \right\},~{\cal Q}_2 = \left\{ X \in \symm^n : \Tr X = 1,~X\succeq 0 \right\}\\
{ }\\
f(U) :=  \max_{X \in {\cal Q}_2} \langle TU,X
\rangle - \hat{\phi}(X) ,\ \ \mbox{with }T = I_{n^2}, ~ \hat{\phi}(X) = -\Tr (AX) .
\EA\]
As in \cite{Nest03}, to ${\cal Q}_1$ and ${\cal Q}_2$ we associate norms and so-called prox-functions. To ${\cal Q}_1$, we associate the Frobenius norm in $\symm^n$, and a prox-function defined for $U \in {\cal Q}_1$ by:
\[
d_1(U) = \frac{1}{2} U^TU .
\]
With this choice, the center $U_0$ of ${\cal Q}_1$, defined as:
\[
U_0 := \arg\min_{U \in {\cal Q}_1} \: d_1(U),
\]
is $U_0= 0$, and satisfies $d_1(U_0) = 0$. Moreover, we have:
\[
D_1 := \max_{U \in {\cal Q}_1} \: d_1(U) = n^2 /2 .
\]
Furthermore, the function $d_1$ is strongly convex on its domain, with convexity parameter of $\sigma_1 = 1$ with respect to the Frobenius norm. Next, for ${\cal Q}_2$ we use the dual of the standard matrix norm (denoted $\|\cdot\|^\ast_2$), and a prox-function
\[
d_2(X) = \Tr (X \log X ) + \log (n) ,
\]
where $\log X$ refers to the {\em matrix} (and not componentwise) logarithm, obtained by replacing the eigenvalues of $X$ by their logarithm. The center of the set ${\cal Q}_2$ is $X_0 =n^{-1}I_n$, where $d_2(X_0) = 0$. We have
\[
\max_{X \in {\cal Q}_2} \: d_2(X) \leq \log n := D_2.
\]
The convexity parameter of $d_2$ with respect to $\|\cdot\|_2^\ast$, is bounded below by $\sigma_2 = 1$.  (This non-trivial result is proved in \cite{Nest04a}.) 

Next we compute the $(1,2)$ norm of the operator $T$ introduced above, which is defined as:
\begin{eqnarray*}
\|T\|_{1,2} &:=& \max_{X ,U} \: \langle TX,U \rangle ~:~ \|U\|_F = 1, \: \|X\|^\ast_2 = 1 \\
&=& \max_X \: \|X\|_2 : \|X\|_F \leq 1 \\
&=& 1 .
\end{eqnarray*}

To summarize, the parameters defined above are set as follows:
\[
D_1 = n^2 /2 ,~  \sigma_1 = 1, ~ D_2  = \log(n), ~\sigma_2 = 1, ~ \|T\|_{1,2} = 1.
\]
Let us now explicitly formulate how the regularization and smooth minimization techniques can be applied to the variance maximization problem in (\ref{eq:orig-pb}).

\subsubsection{Regularization} The method in \cite{Nest03} first sets a regularization parameter
\[
\mu := \frac{\epsilon}{2D_2}.
\]
The method then produces an $\epsilon$-suboptimal optimal value and corresponding suboptimal solution in a number of steps not exceeding
\[
N = \frac{4 \|T\|_{1,2}}{\epsilon}
\sqrt{\displaystyle\frac{D_1D_2}{\sigma_1\sigma_2} }.
\]
The non-smooth objective $f(X)$ of the original problem is replaced with
\[
\min_{U \in {\cal Q}_1} \: f_\mu(U) ,
\]
where $f_\mu$ is the penalized function involving the prox-function $d_2$:
\[
f_\mu(U) := \max_{X \in {\cal Q}_2} \langle TU,X \rangle -
\hat{\phi}(X) - \mu d_2(X) .
\]
Note that in our case, the function $f_\mu$ and its gradient are readily computed; see below. The function $f_\mu$ is a smooth uniform approximation to $f$ everywhere on ${\cal Q}_2$, with maximal error $\mu D_2 = \epsilon/2$. Furthermore, $f_\mu$ has a  Lipschitz continuous gradient, with Lipschitz constant given by:
\[
L := \frac{D_2\|T\|_{1,2}^2}{\epsilon 2 \sigma_2} .
\]
In our specific case, the function $f_\mu$ can be computed explicitly as:
\[
f_\mu(U) = \mu \log \left(\Tr \exp((A+U)/\mu)\right) - \mu \log n,
\]
which can be seen as a smooth approximation to the function $f(U) = \lambdamax(A+U)$. This function $f_\mu$ has a Lipshitz-continuous gradient and is a uniform approximation of the function $f$.

\subsubsection{First-order minimization}
An optimal gradient algorithm for minimizing convex functions with Lipschitz continuous gradients is then applied to the smooth convex function $f_\mu$ defined above. The key difference between the minimization scheme developed in \cite{Nest83} and classical gradient minimization methods is that it is not a descent method but achieves a complexity of $O(L/k^2)$ instead of $O(1/k)$ for gradient descent, where $k$ is the number of iterations and $L$ the Lipschitz constant of the gradient. Furthermore, this convergence rate is provably optimal for this particular class of convex minimization problems (see \cite[Th.~2.1.13]{Nest03a}). Thus, by sacrificing the (local) properties of descent directions, we improve the (global) complexity estimate by an order of magnitude.

For our problem here, once the regularization parameter $\mu$ is set, the algorithm proceeds as follows. 
\vskip .1in
\noindent {\bf Repeat:}
\begin{enumerate}
\item Compute $f_\mu(U_k)$ and $\nabla f_\mu(U_k)$
\item Find $Y_k = \arg\min_{Y \in {\cal Q}_1} \: \langle \nabla f_\mu(U_k) , Y \rangle + \frac{1}{2} L \|U_k-Y\|_F^2$
\item Find $W_k = \arg\min_{W\in {\cal Q}_1} \left\{ \frac{L d_1(W)}{\sigma_1}  + \sum_{i=0}^k \frac{i+1}{2}(f_\mu(U_i)+\langle \nabla f_\mu(U_i),W-U_i \rangle ) \right\}$
\item Set $U_{k+1} = \frac{2}{k+3} W_k + \frac{k+1}{k+3} Y_k$
\end{enumerate}
\noindent {\bf Until} gap $\leq\epsilon$.
\vskip .1in

Step one above computes the (smooth) function value and gradient. The second step computes the \emph{gradient mapping}, which matches the gradient step for unconstrained problems (see \cite[p.86]{Nest03a}). Step three and four update an \emph{estimate sequence} see (\cite[p.72]{Nest03a}) of $f_\mu$ whose minimum can be computed explicitly and gives an increasingly tight upper bound on the minimum of $f_\mu$. We now present these steps in detail for our problem (we write $U$ for $U_k$ and $X$ for $X_k$).

\paragraph{Step 1}
The most expensive step in the algorithm is the first, the computation of $f_\mu$ and its gradient. Setting $Z = A+U$, the problem boils down to computing
\begin{equation}\label{eq:ustar-def}
u^\ast(z) : = \arg\max_{X \in {\cal Q}_2} \langle Z,X \rangle - \mu d_2(X)
\end{equation}
and the associated optimal value $f_\mu(U)$. It turns out that this problem has a very simple solution, requiring only an eigenvalue decomposition for $Z=A+U$. The gradient of the objective function with respect to $Z$ is set to the maximizer $u^\ast(Z)$ itself, so the gradient with respect to $U$ is $\nabla f_\mu(U) = u^\ast(A+U)$.

To compute $u^\ast(Z)$,  we form an eigenvalue decomposition $Z = VDV^T$, with $D = \diag(d)$ the matrix with diagonal $d$, then set
\[
h_i := \frac{\exp( \frac{d_i-d_{\rm max}}{\mu})}{\sum_{j=1}^n \exp (\frac{d_j-d_{\rm max}}{\mu} ) },\quad i=1,\ldots,n,
\]
where $d_{\rm max}:= \max_{\{j=1,\ldots,n\}} d_j$ is used to mitigate problems with large numbers. We then let $u^\ast(z) = VHV^T$, with $H = \diag(h)$. The corresponding function value is given by:
\[
f_\mu(U) = \mu\log\left( \Tr  \exp\left(\frac{(A+U)}{\mu}\right)\right) = \mu\log \left(
\sum_{i=1}^n \exp\left( \frac{d_i}{\mu} \right) \right) - \mu \log n,
\]
which can be reliably computed as:
\[
f_\mu(U) = d_{\rm max} + \mu \log \left( \sum_{i=1}^n \exp(\frac{d_i-d_{\rm max}}{\mu}) \right) - \mu \log n .
\]

\paragraph{Step 2}
This step involves a problem of the form:
\[
\arg\min_{Y \in {\cal Q}_1} \: \langle \nabla f_\mu(U) , Y \rangle + \frac{1}{2} L \|U-Y\|_F^2 ,
\]
where $U$ is given.  The above problem can be reduced to a Euclidean projection:
\begin{equation}\label{eq:euclo-prox}
\arg\min_{\|Y\|_\infty \leq 1} \: \|Y - V\|_F,
\end{equation}
where $V = U - L^{-1}\nabla f_\mu(U)$ is given. The solution is given by:
\[
Y_{ij} = \mbox{\bf sgn}(V_{ij}) \min(|V_{ij}|,1), \quad i,j=1,\ldots,n.
\]

\paragraph{Step 3} The third step involves solving a Euclidean projection problem similar to (\ref{eq:euclo-prox}), with $V$ defined by:
\[
V = -\frac{\sigma_1}{L} \sum_{i=0}^k \frac{i+1}{2} \nabla f_\mu(U_i).
\]

\paragraph{Stopping criterion}  We can stop the algorithm when the duality gap is smaller than $\epsilon$:
\[
\mathrm{gap}_k=\lambdamax(A+U_k) - \Tr AX_k + \ones^T |X_k|\ones \leq \epsilon,
\]
where $X_k = u^\ast((A+U_k)/\mu)$ is our current estimate of the dual variable (the function $u^\ast$ is defined by (\ref{eq:ustar-def})).  The above gap is necessarily non-negative, since both $X_k$ and $U_k$ are feasible for the primal and dual problem, respectively. This is checked periodically, for example every $100$ iterations.

\paragraph{Complexity}  Since each iteration of the algorithm requires computing a matrix exponential (which requires an eigenvalue decomposition and $O(n^3)$ flops in our code), the predicted worst-case complexity to achieve an objective with absolute accuracy less than $\epsilon$ is \cite{Nest03}:
\[
4 \|T\|_{1,2} \frac{O(n^3)}{\epsilon}\sqrt{\displaystyle\frac{D_1D_2}{\sigma_1\sigma_2} } = O(n^4 \sqrt{\log n} / \epsilon).
\]
In some cases, this complexity estimate can be improved by using  specialized algorithms for computing the matrix exponential (see \cite{Mole03} for a discussion). For example, computing only a few eigenvalues might be sufficient to obtain this exponential with the required precision (see \cite{dAsp05}). In our preliminary experiments, the standard technique using Pad\'e approximations, implemented in packages such as Expokit (see \cite{Sidj98}), required too much scaling to be competitive with a full eigenvalue decomposition.

\section{Numerical results \& Applications}
\label{sec:numerical-results} In this section, we illustrate the effectiveness of the proposed approach (called DSPCA in what follows) both on an artificial data set and a real-life data set. We compare with the other approaches mentioned in the introduction: PCA, PCA with simple thresholding, SCoTLASS and SPCA. The results show that our approach can achieve more sparsity in the principal components than SPCA (the current state-of-the-art method) does, while explaining as much variance. The other approaches can explain more variance, but result in principal components that are far from sparse. We begin by a simple example illustrating the link between $k$ and the cardinality of the solution.

\subsection{Artificial data}
To compare the numerical performance with that of existing algorithms, we consider the simulation example proposed by \cite{Zou04}. In this example, three hidden factors are created:
\[
V_1 \sim \mathcal{N}(0,290), \ \ V_2 \sim \mathcal{N}(0,300), \ \
V_3 = -0.3V_1 + 0.925 V_2 + \epsilon, \ \ \epsilon \sim
\mathcal{N}(0,300)
\]
with $V_1, V_2$ and $\epsilon$ independent. Afterward, 10
observed variables are generated as follows:
\[
X_i = V_j + \epsilon_i^j, \quad \epsilon_i^j \sim \mathcal{N}(0,1), 
\]
with $j=1$ for $i=1,\ldots,4$, $j=2$ for $i=5,\ldots,8$ and $j=3$ for $i=9,10$ and $\epsilon_i^j$ independent for $j=1,2,3$, $i=1,\ldots,10$. We use the exact covariance matrix to compute principal components using the different approaches.

Since the three underlying factors have roughly the same variance, and the first two are associated with four variables while the last one is associated with only two variables, $V_1$ and $V_2$ are almost equally important, and they are both significantly more
important than $V_3$. This, together with the fact that the first two principal components explain more than $99\%$ of the total variance, suggests that considering two sparse linear combinations of the original variables should be sufficient to explain most of the variance in data sampled from this model \cite{Zou04}. The ideal solution would thus be to use only the variables $(X_1, X_2, X_3, X_4)$ for the first sparse principal component, to recover the factor $V_1$, and only $(X_5,X_6, X_7, X_8)$ for the second sparse principal component to recover $V_2$.

Using the true covariance matrix and the oracle knowledge that the ideal sparsity is four, \cite{Zou04} performed SPCA (with $\lambda = 0$). We carry out our algorithm with $k = 4$. The results are reported in Table \ref{table:data_art}, together with results for PCA, simple thresholding and SCoTLASS ($t = 2$). Notice that DSPCA, SPCA and SCoTLASS all find the correct sparse principal components, while simple thresholding yields inferior performance. The latter wrongly includes the variables $X_9$ and $X_{10}$ (likely being misled by the high correlation between $V_2$ and $V_3$), moreover, it assigns higher loadings to $X_9$ and $X_{10}$ than to each of the variables $(X_5, X_6, X_7, X_8)$ that are clearly more important. Simple thresholding correctly identifies the second sparse principal component, probably because $V_1$ has a lower correlation with $V_3$. Simple thresholding also explains a bit less variance than the other methods.

\begin{table}[hpt]
\caption{Loadings and explained variance for the first two principal components of the artificial example. Here, ``ST'' denotes the simple thresholding method, ``other'' is all the other methods: SPCA, DSPCA and SCoTLASS. PC1 and PC2 denote the first and second principal components.}\label{table:data_art}
\begin{center}
\scriptsize{
\tabcolsep .03in
\begin{tabular}{c|rrrrrrrrrrc}
& $X_1$ & $X_2$ & $X_3$ & $X_4$ & $X_5$ & $X_6$ & $X_7$& $X_8$ & $X_9$ & $X_{10}$ & explained variance \\
\hline PCA, PC1 & .116 & .116 & .116 & .116 & -.395 &
-.395 & -.395 & -.395 & -.401 & -.401 & $60.0 \%$ \\
PCA, PC2 & -.478 & -.478 & -.478 & -.478 & -.145 & -.145 &
-.145 & -.145 & .010 & .010 & $39.6 \%$ \\
\hline ST, PC1 & 0 & 0 & 0 & 0 & 0 & 0 &
-.497 & -.497 & -.503 & -.503 & $38.8 \%$ \\
ST, PC2 & -.5 & -.5 & -.5 & -.5 & 0 & 0 &
0 & 0 & 0 & 0 & $38.6 \%$ \\
\hline other, PC1 & 0 & 0 & 0 & 0 & .5 & .5 &
.5 & .5 & 0 & 0 & $40.9 \%$ \\
other, PC2 & .5 & .5 & .5 & .5 & 0 & 0 &
0 & 0 & 0 & 0 & $39.5 \%$ \\
\hline
\end{tabular}}
\end{center}
\end{table}

\subsection{Pit props data}
The pit props data (consisting of 180 observations and 13 measured variables) was introduced by \cite{Jeff67} and is another benchmark example used to test sparse PCA codes. Both SCoTLASS  \cite{Joll03} and SPCA \cite{Zou04} have been tested on this data set. As reported in \cite{Zou04}, SPCA performs better than SCoTLASS in the sense that it identifies principal components with 7, 4, 4, 1, 1, and 1 non-zero loadings, respectively, as shown in Table \ref{table:data_pitprop}. This is much sparser than the modified principal components by SCoTLASS, while explaining nearly the same variance ($75.8 \%$ versus $78.2 \%$ for the 6 first principal components) \cite{Zou04}. Also, simple thresholding of PCA, with a number of non-zero loadings that matches the result of SPCA, does worse than SPCA in terms of explained variance.

\begin{figure}[h]
\begin{center}
\begin{tabular}{ccc}
\psfrag{npc}[t][b]{\footnotesize{Number of principal components}}
\psfrag{card}[b][t]{\footnotesize{Cumulative cardinality}}
\psfrag{spca}{\tiny{SPCA}}
\psfrag{k5}{\tiny{$k=5$}}
\psfrag{k6}{\tiny{$k=6$}}
\includegraphics[width=.47\textwidth]{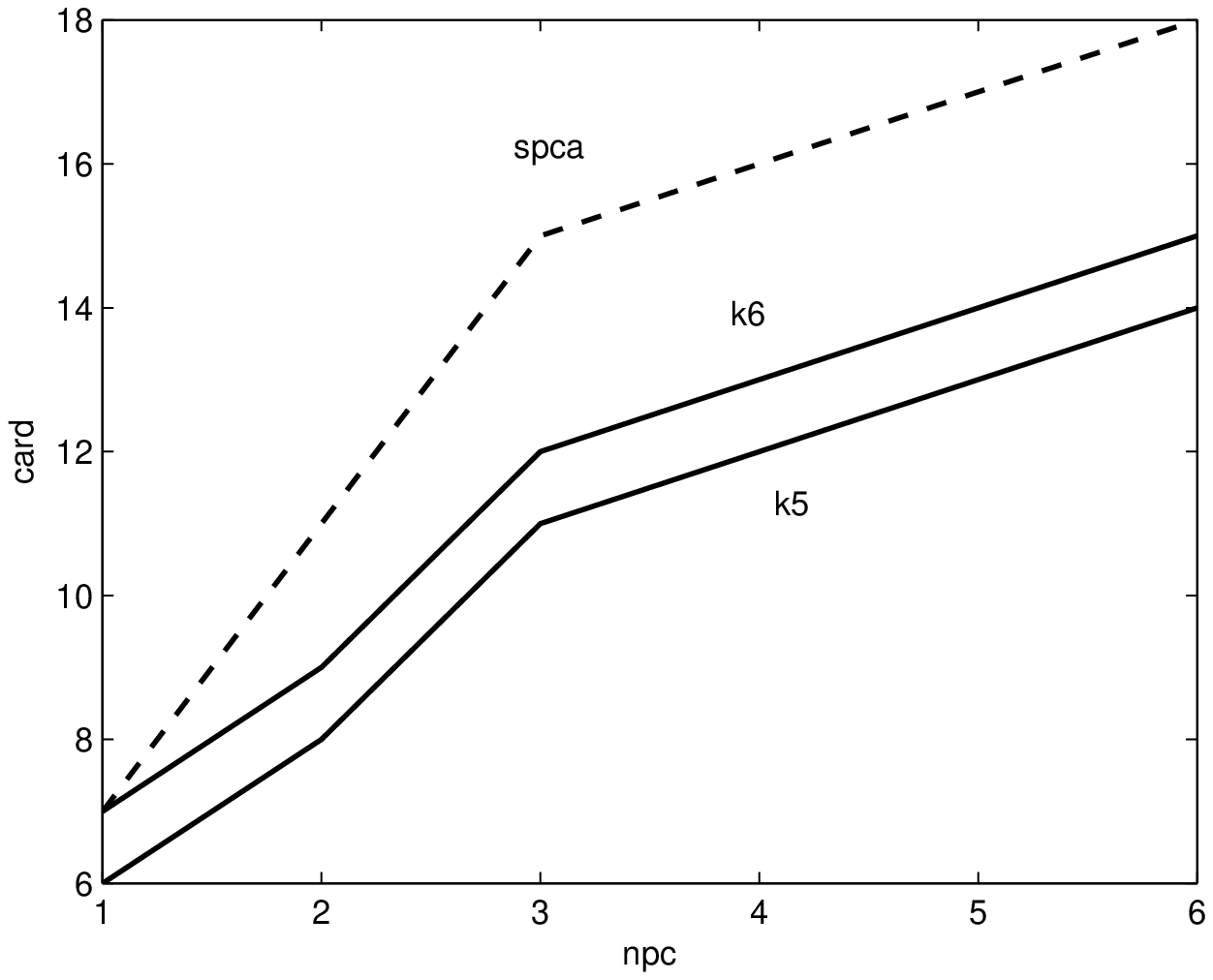} & { } &
\psfrag{npc}[t][b]{\footnotesize{Number of principal components}}
\psfrag{cumvar}[b][t]{\footnotesize{Percentage var. explained}}
\psfrag{pca}{\tiny{PCA}}
\includegraphics[width=.47\textwidth]{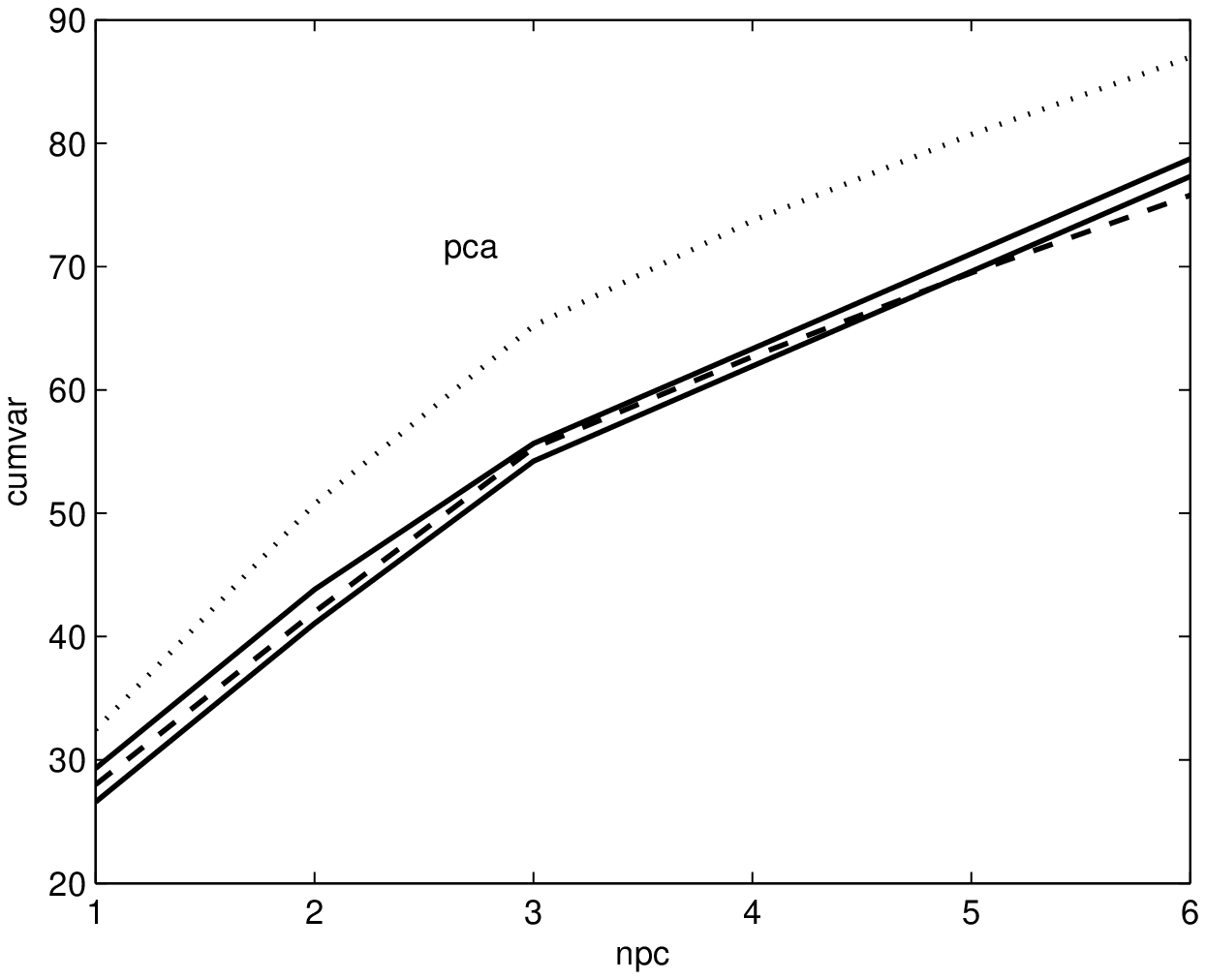} \\
\end{tabular}
\caption{Cumulative cardinality and percentage of total variance explained versus number of principal components, for SPCA and DSPCA on the pit props data. The dashed lines are SPCA and the solid ones are DSPCA with $k_1 = 5$ and $k_1 = 6$. On the right, the dotted line also shows the percentage of variance explained by standard (non sparse) PCA. While explaining the same cumulative variance, our method (DSPCA) produces sparser factors.
\label{fig:pitprop}}
\end{center}
\end{figure}

Following this previous work, we also consider the first 6
principal components.  We try to identify principal components
that are sparser than those of SPCA, but explain the same variance.  Therefore, we choose values for $k$ of 5, 2, 2, 1, 1, 1 (two less than the values of SPCA reported above, but no less than 1). Figure \ref{fig:pitprop} shows the cumulative number of non-zero loadings and the cumulative explained variance (measuring the variance in the subspace spanned by the first $i$ eigenvectors).  It can be seen that our approach is able to explain nearly the same variance as the SPCA method, while clearly reducing the number of non-zero loadings for the first six principal components. Adjusting the first value of $k$ from 5 to 6 (relaxing the sparsity), we obtain results that are still better in terms of sparsity, but with a cumulative explained variance that is uniformly larger than SPCA. Moreover, as in the SPCA approach, the important variables associated with the six principal components do not overlap, which leads to a clearer interpretation. Table \ref{table:data_pitprop} shows the first three corresponding principal components for the different approaches (DSPCAw5 denotes runs with $k_1 = 5$ and DSPCAw6 uses $k_1 = 6$).

\begin{table}[ht]
\caption{Loadings for first three principal components, for the pit props data. DSPCAw5 (resp. DSPCAw6) shows the results for our technique with $k_1$ equal to 5 (resp. 6).} \label{table:data_pitprop}
\begin{center}
\scriptsize{
\tabcolsep .03in
\begin{tabular}{c|rrrrrrrrrrrrr}
& topd & length & moist & testsg & ovensg & ringt & ringb & bowm & bowd & whorls & clear & knots & diaknot \\
\hline SPCA 1  & -.477 & -.476 & 0 & 0 & .177 & 0 & -.250 & -.344 & -.416 & -.400 & 0 & 0 & 0 \\
SPCA 2 & 0 & 0 & .785 & .620 & 0 & 0 & 0 & -.021 & 0 & 0 & 0 & .013 & 0 \\
SPCA 3 & 0 & 0 & 0 & 0 & .640 & .589 & .492 & 0 & 0 & 0 & 0 & 0 & -.015 \\
\hline DSPCAw5 1 & -.560 & -.583 & 0 & 0 & 0 & 0 & -.263 & -.099 & -.371 & -.362 & 0 & 0 & 0 \\
DSPCAw5 2 & 0 & 0 & .707 & .707 & 0 & 0 & 0 & 0 & 0 & 0 & 0 & 0 & 0 \\
DSPCAw5 3 & 0 & 0 & 0 & 0 & 0 & -.793 & -.610 & 0 & 0 & 0 & 0 & 0 & .012 \\
\hline DSPCAw6 1 & -.491 & -.507 & 0 & 0 & 0 & -.067 & -.357 & -.234 & -.387 & -.409 & 0 & 0 & 0 \\
DSPCAw6 2 & 0 & 0 & .707 & .707 & 0 & 0 & 0 & 0 & 0 & 0 & 0 & 0 & 0 \\
DSPCAw6 3 & 0 & 0 & 0 & 0 & 0 & -.873 & -.484 & 0 & 0 & 0 & 0 & 0 & .057 \\
\hline
\end{tabular}}
\end{center}
\end{table}

\subsection{Controlling sparsity with $k$}
\begin{figure}[htp]
\begin{center}
\begin{tabular}{cc}
\psfrag{k}[t][b]{$k$}
\psfrag{cardinality}[b][t]{Cardinality}
\includegraphics[width=0.50 \textwidth]{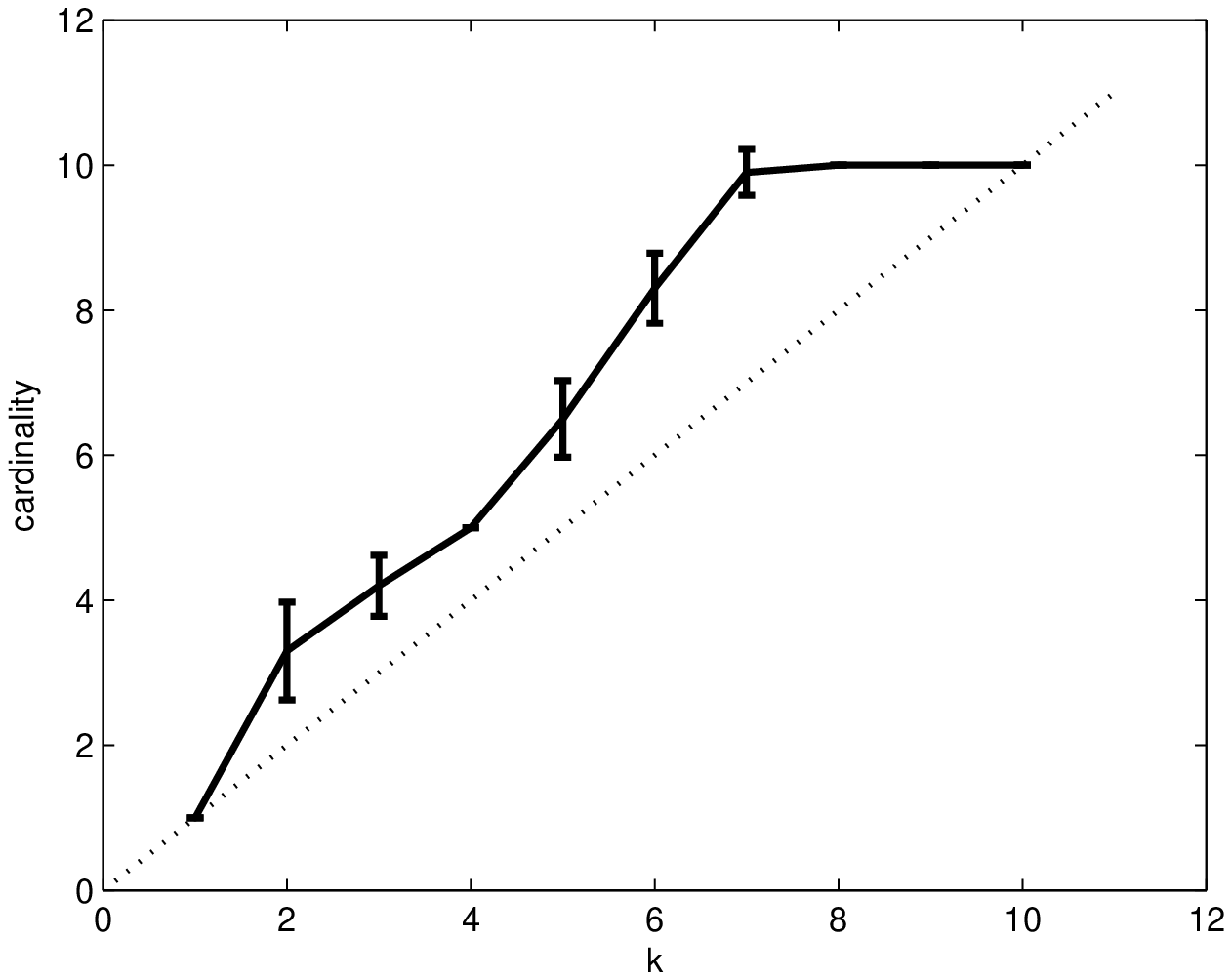}
\psfrag{nsize}[t][b]{Problem size $n$}
\psfrag{cputime}[b][t]{CPU time}
\includegraphics[width=0.50 \textwidth]{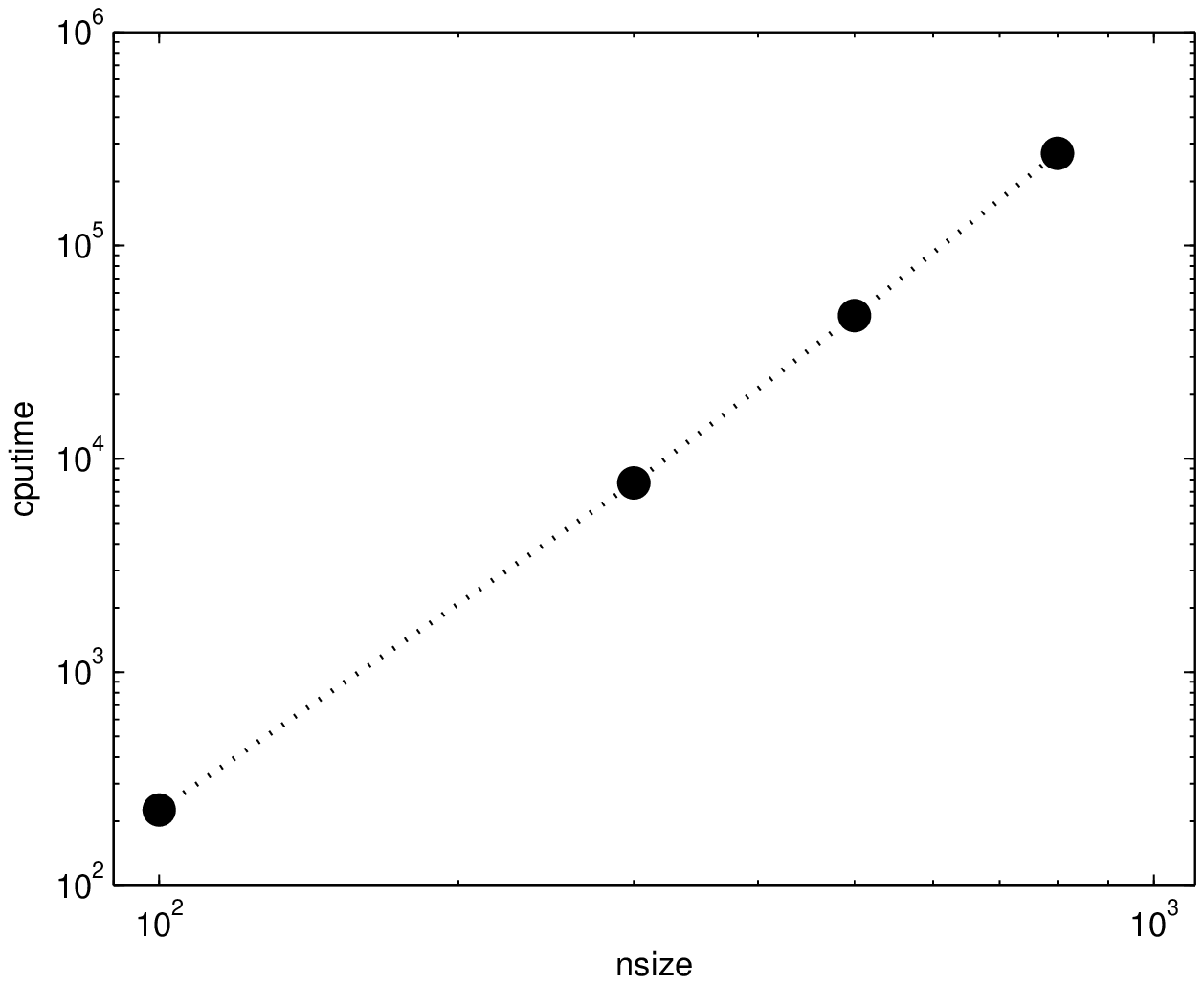}
\end{tabular}
\caption{Left: Plot of average cardinality (and its standard deviation) versus $k$ for 100 random examples with original cardinality 5. Right: Plot of CPU time (in seconds) versus problem size for randomly chosen problems. \label{fig:card-versus-k}}
\end{center}
\end{figure}

We present a simple example to illustrate how the sparsity of the
solution to our relaxation evolves as $k$ varies from $1$ to $n$.
We generate a $10\times 10$ matrix $U$ with uniformly distributed
coefficients in $[0,1]$. We let $v$ be a sparse vector with:
\[
v=(1,0,1,0,1,0,1,0,1,0).
\]
We then form a test matrix $A=U^TU+\sigma vv^T$, where $\sigma$ is a signal-to-noise ratio that we set equal to $15$. We sample $50$ different matrices $A$ using this technique. For each value of $k$ between $1$ and $10$ and each $A$, we solve the following SDP:
\[
\BA{ll}
\mbox{max} & \Tr(AX)\\
\mbox{subject to} & \Tr(X)=1\\
 & \ones ^T |X| \ones \leq k\\
 & X \succeq 0.
\EA
\]
We then extract the first eigenvector of the solution $X$ and record its cardinality. In Figure \ref{fig:card-versus-k}, we show the mean cardinality (and standard deviation) as a function of $k$. We observe that $k+1$ is actually a good predictor of the cardinality, especially when $k+1$ is close to the actual cardinality ($5$ in this case). In fact, in the random examples tested here, we always recover the original cardinality of 5 when $k+1$ is set to 5.

\subsection{Computing Time versus Problem Size}
In Figure \ref{fig:card-versus-k} we plot the total CPU time used for randomly chosen problems of size $n$ ranging from 100 to 800. The required precision was set to $\epsilon=10^{-3}$, which was always reached in fewer than 60000 iterations. In these examples, the empirical complexity appears to grow as $O(n^3)$.

\subsection{Sparse PCA for Gene Expression Data Analysis}
We are given $m$ data vectors $x_j\in\reals^n$, with $n=500$. Each coefficient $x_{ij}$ corresponds to the expression of gene $i$ in experiment $j$. For each vector $x_j$ we are also given a class $c_j\in\{0,1,2,3\}$. We form $A=xx^T$, the covariance matrix of the experiment. Our objective is to use PCA to first reduce the dimensionality of the problem and then look for \emph{clustering} when the data are represented in the basis formed by the first three principal components. Here, we do not apply any clustering algorithm to the data points, we just assign a color to each sample point in the three dimensional scatter plot, based on known experimental data.

\begin{figure}[htp]
\begin{center}
\psfrag{f1}{$f_1$}
\psfrag{f2}{$f_2$}
\psfrag{f3}{$f_3$}
\psfrag{g1}{$g_1$}
\psfrag{g2}{$g_2$}
\psfrag{g3}{$g_3$}
\psfrag{PCA}[b][c]{PCA}
\psfrag{DSPCA}[b][c]{Sparse PCA}
\includegraphics[width=\linewidth]{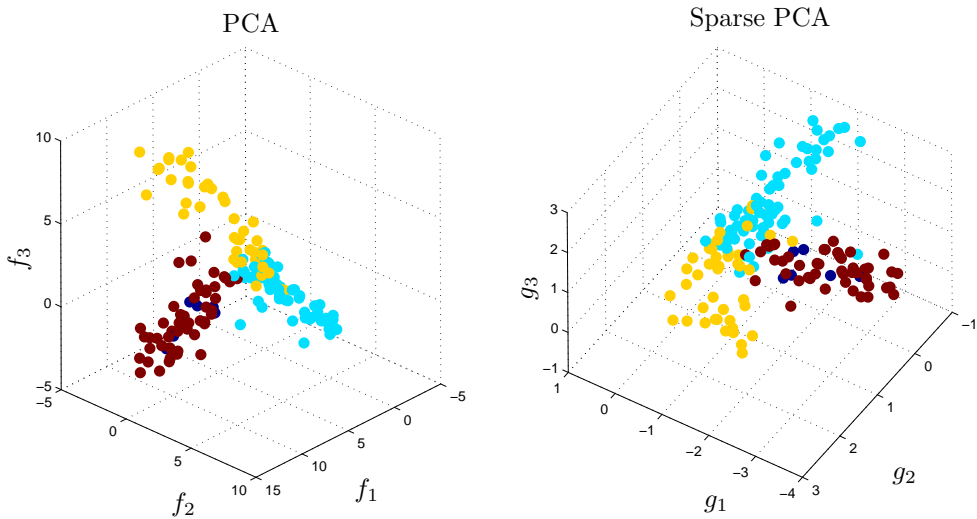}
\caption{\label{fig:DSPCA-genes} Clustering of the gene expression data in the PCA versus sparse PCA basis with 500 genes. The factors $f$ on the left are dense and each use all 500 genes while the sparse factors $g_1,~g_2$ and $g_3$ on the right involve 6, 4 and 4 genes respectively. (Data: Iconix Pharmaceuticals)}
\end{center}
\end{figure}

The sparsity of the factors in sparse PCA implies that the clustering can be attributed to fewer genes, making interpretation easier. In Figure {\ref{fig:DSPCA-genes}}, we see clustering in the PCA representation of the data and in the DSPCA representation.  Although there is a slight drop in the resolution of the clusters for DSPCA, the key feature here is that the total number of nonzero gene coefficients in the DSPCA factors is equal to 14 while standard PCA produces three dense factors, each with 500 nonzero coefficients.

\section*{Acknowledgments}
Thanks to Andrew Mullhaupt and Francis Bach for useful
suggestions. We would like to acknowledge support from NSF grant 0412995, ONR MURI N00014-00-1-0637, Eurocontrol-C20052E/BM/03 and C20083E/BM/05, NASA-NCC2-1428 and a gift from Google, Inc.

\small{
\bibliographystyle{siam}
\bibliography{MainPerso}}
\end{document}